%% file: IEEE-conference-template-062824.tex
\documentclass[letterpaper, 10 pt, conference]{ieeeconf}  

\IEEEoverridecommandlockouts                              

\usepackage{cite}
\usepackage{amsmath,amssymb,amsfonts}
\usepackage{algorithm}
\usepackage{algorithmic}
\usepackage{graphicx}
\usepackage{textcomp}
\usepackage{xcolor}
\usepackage{listings}
\usepackage{tabularx}
\usepackage{siunitx}
\usepackage{booktabs}
\usepackage{caption}
\usepackage{comment}
\usepackage{balance}
\usepackage{array}
\usepackage{balance}
\usepackage{subcaption}

\usepackage{acro}

\DeclareAcronym{EMF}{
	short = EMF,
	long  = Eclipse Modeling Framework,
	sort  = abbrev,
}

\DeclareAcronym{UML}{
	short = UML,
	long  = Unified Modeling Language,
	sort  = abbrev,
}

\DeclareAcronym{SysML}{
	short = SysML,
	long  = Systems Modeling Language,
	sort  = abbrev,
}

\DeclareAcronym{MBE}{
	short = MBE,
	long  = Model-based Engineering,
	sort  = abbrev,
}

\DeclareAcronym{LLM}{
	short = LLM,
	long  = Large Language Model,
	sort  = abbrev,
}

\DeclareAcronym{OMG}{
	short = OMG,
	long  = Object Management Group,
	sort  = abbrev,
}

\DeclareAcronym{MOF}{
	short = MOF,
	long  = Meta-Object Facility,
	sort  = abbrev,
}

\definecolor{darkgreen}{rgb}{0.0, 0.39, 0.0}

\begin{document}
	
	\title{Automating Automotive Software Development: \\A Synergy of Generative AI and Model-Based Methods}
	
	\author{Fengjunjie Pan, Yinglei Song, Long Wen, Nenad Petrovic, Krzysztof Lebioda and Alois Knoll
		\thanks{F. Pan,  Y. Song, L. Wen, N. Petrovic, K. Lebioda, A. Knoll are with Robotics, Artificial Intelligence and Real-Time Systems, School of Computation, Information and Technology, Technical University of Munich, Munich, Germany. \{panf, syin, wenl, pne, lebioda, knoll\}@in.tum.de}%
	}

	\maketitle
	
	\begin{abstract}

As the automotive industry shifts its focus toward software-defined vehicles, the need for faster and reliable software development continues to grow. However, traditional methods show their limitations.
The rise of Generative Artificial Intelligence (GenAI), particularly Large Language Models (LLMs), introduces new opportunities to automate automotive software development tasks such as requirement analysis and code generation. However, due to the complexity of automotive systems, where software components must interact with each other seamlessly, challenges remain in software integration and system-level validation.
In this paper, we propose to combine GenAI with model-driven engineering to automate automotive software development. Our approach uses LLMs to convert free-text requirements into event chain descriptions and to generate platform-independent software components that realize the required functionality. At the same time, formal models are created based on event chain descriptions to support system validation and the generation of integration code for integrating generated software components in the whole vehicle system through middleware.
This approach increases development automation via GenAI while enabling model-based analysis to improve system reliability. We demonstrated our method with different LLMs and tested it in the CARLA simulation environment with ROS2 middleware. We evaluated the system in a simple Autonomous Emergency Braking scenario.
	\end{abstract}

	\section{Introduction}
	\input{sections/01_introduction}

	\section{Related Work} \label{sec:background}
	\input{sections/02_relatedWork}

	\section{Agentic Workflow} \label{sec:proposed approach}
	\input{sections/03_method}

	\section{Experiment} \label{sec:evaluation}
	\input{sections/04_evaluation}

	\section{Conclusion} \label{sec:conclusion}
	\input{sections/05_conclusion}

	\bibliographystyle{IEEEtran}
	\bibliography{ref}
	\balance
 	\section*{Acknowledgment}
 	This research was funded by the Federal Ministry of Research, Technology and Space of Germany as part of the CeCaS project, FKZ: 16ME0800K.
 	
\begin{appendices}
\section{Prompt for event chain generation}\label{lst:promptEventChain}
\begin{lstlisting}[breaklines=true]
# Task description
You are an automotive software system developer. Your goal is to design a structured event chain description in JSON format for the Autonomous Emergency Braking system, using existing components and signals/messages where possible. If necessary, define new software components that adhere to the same format.

# Guidelines for Building the Event Chain
1. Reuse existing components and topics in the event chain if available.
2. The event chain should contains only necessary inputs and outputs of each components.
3. The component sequence in the generated json description is the software sequence in the event chain.

# Template for Event Chain Description
Use the structure below for each component in the event chain description:
[{
 "name": "ComponentName",
 "description": "What the component does. What is the implementation logic based on requirements",
 "input": [
 {
  "topic": "/some/input_topic",
  "message_type": "some_msgs/MessageType",
  "qos_profile": "quality_of_service_profile",
  "values": [
  {
   "name": "input_value_name",
   "field": "actual_field_name",
   "description": "What this input value means"
  }]
 }],
 "output": [
 (follow the same structure as for the input data)
 ]
}]

# Existing software components
{{software component description in json format}}

# Existing signals/messages
[{
 "Topic Name": "/topic_name",
 "Message Type": "MessageType",
 "qos_profile": "quality_of_service_profile",
 "Message Definition": [
 {
  "Field": "field_name",
  "Type": "dataType",
  "Description": "What this value means"
 }]
}]

# Generated Event Chain
\end{lstlisting}
\section{Prompt for function code generation}\label{lst:prompt_code}
\begin{lstlisting}[breaklines=true]
# Task description
You are an automotive software developer responsible for implementing a submodule for the entire system.
The submodule must fulfill the given software description in JSON format.
The submodule is middleware-independent.
A middleware wrapper code (it only passes raw inputs to the submodule and routes outputs from it) can be used to call the submodule and integrate it into the system.

# Instructions
1. The software submodule must be a self-contained, standalone script of a Python class with all dependencies.
2. The submodule functionality should be executed directly in the function execute(input1, input2, ...)
3. The output of function execute(input1, input2, ...) is a dict {'output1':output1, 'output2': output2, ...}
4. The input and output data of execute() should strictly follow the submodule description.
5. Do not include any middleware-specific code.
6. The software submodule must encapsulate all necessary logic, data processing, and state management to fulfill the system functionality.
7. If the software component is stateful, the state must be fully managed internally within the class.
8. The class must be designed to work with raw input values (as passed by the middleware).
9. The class uses the name from the submodule description.
10. Please only implement the submodule, not the function of the entire target system.

# The submodule description
{{Description of target software extracted from the event chain description.}}

# The generated sub software module is:
\end{lstlisting}

\section{Acceleo template for ROS node generation}\label{lst:acceleo}
\begin{lstlisting}[caption={},label={lst:acceleo},breaklines=true]
(*@\textcolor{red}{[template public main(eventchain : EventChain)]} @*)
(*@\textcolor{darkgreen}{[comment @main /]} @*)
(*@\textcolor{purple}{[for (}@*)(*@\textcolor{blue}{node:SoftwareNode|eventchain.software}@*)(*@\textcolor{purple}{)]} @*)
(*@\textcolor{purple}{[file (}@*)(*@\textcolor{blue}{node.name.toLowerCase().concat('\_node.py'), false, }@*)
(*@\textcolor{darkgreen}{UTF-8'}@*)(*@\textcolor{purple}{)]}@*)
... (Import statements omitted for brevity)

class [(*@\textcolor{blue}{node.name.concat('\_node')}@*)(*@\textcolor{purple}{/]}@*)(Node):
def __init__(self):
super().__init__('[(*@\textcolor{blue}{node.name.concat('\_node')}@*)(*@\textcolor{purple}{/]}@*)')
self.[(*@\textcolor{blue}{node.name}@*)(*@\textcolor{purple}{/]}@*) = [(*@\textcolor{blue}{node.name}@*)(*@\textcolor{purple}{/]}@*)()
(*@\textcolor{purple}{[for (}@*)(*@\textcolor{blue}{data : Data | node.input}@*))]
self.(*@\textcolor{purple}{[}@*)(*@\textcolor{blue}{data.name}@*)(*@\textcolor{purple}{/]}@*) = None
(*@\textcolor{purple}{[/for]}@*)
(*@\textcolor{purple}{[for (}@*)(*@\textcolor{blue}{data : Data | node.input}@*))]
self.(*@\textcolor{purple}{[}@*)(*@\textcolor{blue}{data.name}@*)(*@\textcolor{purple}{/]}@*)_subscriber = self.create_subscription((*@\textcolor{purple}{[}@*)(*@\textcolor{blue}{data.messageType.tokenize('/')->last()}@*)(*@\textcolor{purple}{/]}@*), (*@\textcolor{black}{"}@*)(*@\textcolor{purple}{[}@*)(*@\textcolor{blue}{data.topicName}@*)(*@\textcolor{purple}{/]}@*)(*@\textcolor{black}{"}@*), self.(*@\textcolor{purple}{[}@*)(*@\textcolor{blue}{data.name}@*)(*@\textcolor{purple}{/]}@*)_callback, qos_profile=10)
(*@\textcolor{purple}{[/for]}@*)
(*@\textcolor{purple}{[for (}@*)(*@\textcolor{blue}{data : Data | node.output}@*)(*@\textcolor{purple}{)]}@*)
self.(*@\textcolor{purple}{[}@*)(*@\textcolor{blue}{data.name}@*)(*@\textcolor{purple}{/]}@*)_publisher= self.create_publisher((*@\textcolor{purple}{[}@*)(*@\textcolor{blue}{data.messageType.tokenize('/')->last()}@*)(*@\textcolor{purple}{/]}@*), (*@\textcolor{black}{"}@*)(*@\textcolor{purple}{[}@*)(*@\textcolor{blue}{data.topicName}@*)(*@\textcolor{purple}{/]}@*)(*@\textcolor{black}{"}@*), qos_profile=10)
(*@\textcolor{purple}{[/for]}@*)
self.timer = self.create_timer(1.0/(*@\textcolor{purple}{[}@*)(*@\textcolor{blue}{node.frequency}@*)(*@\textcolor{purple}{/]}@*), self.execute)

(*@\textcolor{purple}{[for (}@*)(*@\textcolor{blue}{data : Data | node.input}@*))]
def (*@\textcolor{purple}{[}@*)(*@\textcolor{blue}{data.name}@*)(*@\textcolor{purple}{/]}@*)_callback(self, data):
self.(*@\textcolor{purple}{[}@*)(*@\textcolor{blue}{data.name}@*)(*@\textcolor{purple}{/]}@*) = data.(*@\textcolor{purple}{[}@*)(*@\textcolor{blue}{data.fieldName}@*)(*@\textcolor{purple}{/]}@*)
(*@\textcolor{purple}{[/for]}@*)

def execute(self):
(*@\textcolor{purple}{[for (}@*)(*@\textcolor{blue}{data : Data | node.input}@*))]
if self.(*@\textcolor{purple}{[}@*)(*@\textcolor{blue}{data.name}@*)(*@\textcolor{purple}{/]}@*) is None:
self.get_logger().warn((*@\textcolor{black}{"}@*)msg not received(*@\textcolor{black}{"}@*))
return
(*@\textcolor{purple}{[/for]}@*)
output = self.[(*@\textcolor{blue}{node.name}@*)(*@\textcolor{purple}{/]}@*).execute((*@\textcolor{purple}{[for (}@*)(*@\textcolor{blue}{data : Data | node.input}@*))](*@\textcolor{purple}{[}@*)(*@\textcolor{blue}{data.name}@*)(*@\textcolor{purple}{/]}@*)=self.(*@\textcolor{purple}{[}@*)(*@\textcolor{blue}{data.name}@*)(*@\textcolor{purple}{/]}@*)(*@\textcolor{purple}{[if}@*) (*@\textcolor{blue}{(node.input->indexOf(data) <> node.input->size()}@*)(*@\textcolor{purple}{)]}@*), (*@\textcolor{purple}{[/if]}@*)(*@\textcolor{purple}{[/for]}@*))
(*@\textcolor{purple}{[for (}@*)(*@\textcolor{blue}{data : Data | node.output}@*)(*@\textcolor{purple}{)]}@*)
(*@\textcolor{purple}{[}@*)(*@\textcolor{blue}{data.name}@*)(*@\textcolor{purple}{/]}@*)_msg = (*@\textcolor{purple}{[}@*)(*@\textcolor{blue}{data.messageType.tokenize('/')->last()}@*)(*@\textcolor{purple}{/]}@*)()
(*@\textcolor{purple}{[}@*)(*@\textcolor{blue}{data.name}@*)(*@\textcolor{purple}{/]}@*)_msg.(*@\textcolor{purple}{[}@*)(*@\textcolor{blue}{data.fieldName}@*)(*@\textcolor{purple}{/]}@*) = output['(*@\textcolor{blue}{data.name}@*)'(*@\textcolor{purple}{/]}@*)
self.(*@\textcolor{purple}{[}@*)(*@\textcolor{blue}{data.name}@*)(*@\textcolor{purple}{/]}@*)_publisher.publish((*@\textcolor{purple}{[}@*)(*@\textcolor{blue}{data.name}@*)(*@\textcolor{purple}{/]}@*)_msg)
(*@\textcolor{purple}{[/for]}@*)
... (ROS main function definition omitted for brevity)
(*@\textcolor{purple}{[/file]}@*)
(*@\textcolor{purple}{[/for]}@*)
(*@\textcolor{red}{[/template]}@*)
\end{lstlisting}

\end{appendices}
\end{document}

%% file: sections/01_introduction.tex
Over the past decade, the automotive industry has been undergoing a transformation from the traditional vehicles focused on mechanical systems to software-defined vehicles~(SDVs).
As the amount of software in vehicles increases, the complexity of automotive software development is growing rapidly. 
Modern cars are now estimated to be about five times more complex than they were just a few years ago~\cite{mckinsey}. 
Meanwhile, productivity in software development is increasing only slowly, which creates a widening gap between rising demands and the available development capacity.

Recently, the emergence of Generative Artificial Intelligence (GenAI), particularly Large Language Models (LLMs), has provided new possibilities for automating tasks across various domains, thanks to their strong capabilities in text understanding and generation~\cite{matarazzo2025survey}.
Their potential is now being explored in the automotive sector as well~\cite{phatale2024, petrovic2024-3}. 
LLMs show promise for tasks such as requirement analysis, documentation, and even code generation. However, several challenges limit their practical application in the automotive domain.

Automotive systems are highly complex, involving numerous software and hardware components interacting with each other.  
These systems typically follow a layered software architecture, which promotes modularity and scalability through the use of middleware. 
Although LLMs can assist in generating small software modules, creating complete and dependable system-level software remains a significant challenge. 
Furthermore, integrating these LLM-generated components into existing vehicle systems adds another layer of complexity. 
In addition, automotive platforms often require strict guarantees related to safety, real-time performance, and reliability. These constraints are difficult to analyze using purely statistical AI models. 
In such cases, system validation often depends on model-based formal methods and expert review.

In this work, we aim to accelerate automotive software development by combining the generative capabilities of LLMs with the rigor of model-based methods. 
In the automotive domain, system behavior is often described using the event chain concept~\cite{muenzenberger2021event}, which breaks down high-level software functionality into smaller, well-defined processing steps with clearly specified inputs and outputs. Event chains are frequently used not only to guide software implementation but also to support system analysis~\cite{padma2020}.
Building on this concept, we introduce an event chain-driven development approach based in an agentic workflow that coordinates multiple specialized LLMs. Our process begins by using an LLM to extract functional software behaviors from natural language requirements. These behaviors are structured into event chains, where each step is described as a subcomponent, representing a modular unit of functionality.
For each subcomponent, middleware-independent software is generated using a code-specific LLM.
In parallel, a formal representation of the event chain, e.g., in Eclipse Modeling Framework~(EMF) format, is derived via a model-specific LLM. The formal event chain model allows for early-stage system validation and facilitates post-analysis. 
At the same time, the formal model enables model-based code generation to integrate subcomponents into the overall runtime environment. 
The completed software is then deployed to the vehicle platform or a simulation environment for further testing and analysis.

We demonstrate the feasibility of the proposed approach through a basic Autonomous Emergency Braking (AEB) scenario within a ROS2-based simulation environment using the CARLA simulator~\cite{dosovitskiy2017carla}. Several LLMs, including GPT-4o, Gemini 2.5 Pro, and Llama 3.3 70B, are evaluated for their effectiveness in the proposed workflow.

%% file: sections/02_relatedWork.tex
Our literature review focuses on the use of generative AI and model-based formal methods in automotive software development.

Numerous studies have investigated the use of LLMs to generate automotive-related code from textual descriptions.
Abdalla et al.~\cite{abdalla2024} investigated fine-tuning small open-source LLMs to generate low-level vehicle control functions in graphical programming languages for model-based development environments, such as MATLAB Simulink. The resulting models were subsequently used to generate C code via traditional model-based techniques.
Patil et al.~ \cite{patil2024} focused on generating industrial-grade C code for automotive embedded systems using LLMs. Their work supports both high-level and low-level textual specifications, including the formal ANSI/ISO C Specification Language, and integrates formal code verification tools to analyze the generated output.  
Nouri et al.~\cite{nouri2025} proposed an iterative approach for generating safety-critical vehicle functions. They combined Python code generation with the esmini simulation environment and predefined test cases. If the generated code failed the test, the LLM was prompted again with the test results and previous code to trigger a correction cycle.
These works demonstrate that modern LLMs are capable of producing syntactically and semantically valid code for automotive applications. However, they primarily focus on generating isolated functional scripts and do not address the challenges of integrating multiple software components or how these components interact within a complete vehicle system.

In parallel, advanced model-based approaches have been widely used to support system-level integration and analysis in automotive domain. 
Vinoth Kannan~\cite{VinothKannan2021} provided an overview of model-based methods in automotive systems, highlighting how they can improve quality and reduce development time through verification and automated code generation. 
Holtmann et al.~\cite{holtmann2021} presented a comprehensive model-based development pipeline, starting from system architecture in SysML to software design and integration. They proposed automating the transformation from SysML models to AUTOSAR-compliant software models, which enables seamless software integration into the vehicle runtime.
Our previous work~\cite{pan2024sdv} explored the use of formal models to represent vehicle software and hardware for analyzing resource allocation issues. We also utilized model-based code generation to automate deployment-related software.
These studies confirm the strengths of model-based techniques in automotive development. However, a key challenge remains for the usage of model-based method is significant domain expertise required to construct and use formal models.
Recently, several works~\cite{petrovic2025meta, pan2025model, pan2024ocl} have explored how advanced LLMs can be leveraged to generate model components automatically. This opens new opportunities to combine generative AI and formal methods to automate the development of automotive software systems.

Therefore, this paper investigates the integration of generative AI and model-based engineering for automotive software development. We propose an agent-based workflow that utilizes state-of-the-art LLMs for requirement analysis, function-level code generation, and formal model creation. In addition, model-based methods are employed for system-level analysis and software integration. The key benefit of our approach lies in accelerating software development while supporting formal system analysis and validation.

%% file: sections/03_method.tex
\begin{figure*}[t]
	\centering
	\includegraphics[width=\linewidth]{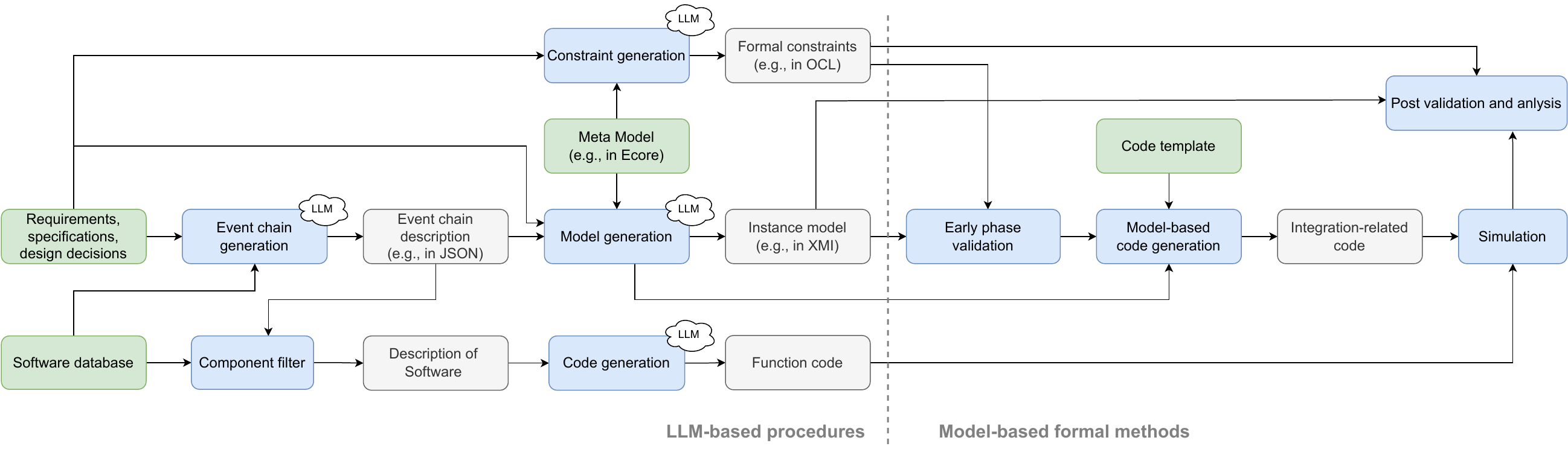}
	\caption{Proposed workflow for automated automotive software development. Inputs to the workflow are highlighted in green, processing steps in blue, and intermediate artifacts in grey.}
	\label{fig:workflow}
\end{figure*}

Our proposed method integrates multiple LLMs and model-based techniques within an agentic workflow. The entire software development process follows an event chain-driven approach. This approach automates the generation of automotive software, facilitates seamless system integration, and enables system-level verification.

The workflow begins with the interpretation of natural language requirements, specifications and design decisions. A general-purpose LLM is used to extract the software behavior from these requirements, which is then translated into an event chain description (Section~\ref{sec:eventgen}). Each step in the event chain represents a subcomponent with explicitly defined inputs and outputs. 

In typical automotive software development, some software components and signals may already exist within the software base of the system. The LLM should consider this context when designing the target software. In the end, the resulting event chain describes both existing and not-yet-implemented subcomponents, along with their interactions through input and output signals. In addition, it captures the implementation logic of each subcomponent as extracted from the requirements (mainly functional requirements, serving as the basis for automated software generation), and, where applicable, includes software properties such as timing or resource constraints (to support model-based system analysis).

Afterwards, the descriptions of not-yet-implemented subcomponents in the event chain are identified and passed to a code-specific LLM for generating platform-independent function code (Section~\ref{sec:LLM-code-gen}). By platform-independent, we refer to code decoupled from specific middleware, operating systems, or hardware dependencies. These generated components adhere to the specifications defined in the event chain. On the other hand, existing components from the software base are reused rather than regenerated.

In parallel, a formal representation of the event chain is constructed using a modeling-specific LLM (Section~\ref{sec:model-gen}). This produces a system model instance conforming to a predefined event chain meta model. The formal model can capture both the event chain's architectural structure and any relevant system properties. In addition, formal constraints will be extracted and formalized from the design decisions, specifications, and requirements (mainly non-functional) manually or through LLMs. These formal models and constraints support early-stage system validation (Section~\ref{sec:OCL}) and facilitates post-simulation analysis by allowing the integration of runtime metrics back into the model. The formal instance model also enables model-based code generation, producing the integration code that connects individual subcomponents into the vehicle runtime system (Section~\ref{sec:model-code-gen}).
The advantage of model-driven integration lies in its analytical approach, which significantly reduces the uncertainty compared to purely statistical approaches driven by LLMs.
As a result, integration can be performed in a deterministic way, allowing testing efforts to concentrate on validating the correctness of the generated software components and their compliance with the functionality defined in requirements, rather than resolving integration errors or interface mismatches.

Finally, the generated software system is deployed to the target platform, which may include hardware-in-the-loop configurations or simulation environments. For example, in our case study, deployment is conducted in the CARLA simulator using ROS 2 middleware.
Simulation-based tests are then executed to verify the system’s functional correctness against the initial requirements. During runtime, behavioral data and system metrics can be collected and inserted into the formal model, enabling further analysis and refinement through iterative development cycles.

\section{LLM-based Generation}
The proposed workflow leverages \acp{LLM} at multiple stages for both requirement analysis and artifact generation. In the first step, an LLM is used to derive a structured event chain description from the provided requirements. Subsequently, the event chain description is modeled into an instance model by an LLM, which then serves as a crucial input for model-based analysis and further development. 
In addition, LLMs are also employed to generate software components that are not available in the existing software base.
\subsection{Event Chain Generation}\label{sec:eventgen}
An event chain captures the data flow within a system and describes how different subcomponents interact with one another. Generating the event chain description is a core enabler of the proposed method for automated automotive software development.

The primary input for this step is a set of requirements that define the expected behavior of the target system. In complex automotive environments, reusing existing software can significantly improve productivity. To support this, information about the existing software base is made available during event chain generation, allowing the system to incorporate previously developed components where applicable.
In addition to software reuse, many vehicle signals are often predefined at runtime. This runtime signal information is also supplied as input for generating the event chain.

The prompt used to guide the LLM in creating the event chain is provided in Appendix~\ref{lst:promptEventChain}.
We utilize a standardized JSON format to structure the relevant software and signal information. This includes component names, textual descriptions, implementation logic, and detailed input/output signal specifications.
As our prototype is implemented using ROS, the signal templates in our examples include ROS-specific fields such as topic names and message types. However, this approach is not limited to ROS. A more generalized signal definition, such as Vehicle Signal Specification~(VSS), can also be adopted. This would require adapters during implementation to convert standardized signal formats into platform-specific implementations. However, the discussion of standardized signal representation and its integration with runtime middleware is beyond the scope of this paper.

The resulting event chain is represented as a list of software components in JSON format. Each component defines its required input/output signal and its implementation logic based on the requirements. This structured event chain description is then passed on to the next stage for software code generation (Section~\ref{sec:LLM-code-gen}) and instance model generation (Section~\ref{sec:model-gen}).

\subsection{Model and Constraints Creation}\label{sec:model-gen}
We follow the model generation approach proposed in \cite{pan2025model} to generate a formal event chain model. This method leverages the capabilities of LLMs to generate a conceptual instance model in JSON format, which is subsequently parsed into a valid instance model file.

This step takes as input a self-defined event chain meta model (Section~\ref{sec:model}) along with the event chain description produced from the previous step (Section~\ref{sec:eventgen}).
In our implementation, the event chain description is provided in a customized JSON format.
Following the strategy outlined in~\cite{pan2025model}, we manually construct a one-shot example comprising the event chain meta model, a similar event chain description, and the expected instance model. This example is used to guide the LLM in generating the conceptual instance model for the target system. The output from the LLM is then parsed by the instance modeler introduced in~\cite{pan2025model}, resulting in a structured instance model that conforms to the predefined meta model. This model serves as a foundation for subsequent tasks, including model-based system analysis and model-driven code generation (Section~\ref{sec:model-code-gen}).

The constraint generation is a relatively straightforward process, as mentioned in \cite{pan2024ocl}. In our case study, we employ the Object Constraint Language (OCL) \cite{OMG2014} as the constraint language for model-based formal constraint description. It has logic similar to first-order logic and is designed for MOF-based model validation. We simply pass the meta model information and natural language text to the LLM, and the corresponding OCL constraints are generated. Examples of generated OCL constraints are presented in Section~\ref{sec:OCL}

\subsection{Code generation}\label{sec:LLM-code-gen}
Code generation is a central focus of this work.
Automotive software, even for individual features such as AEB, is inherently complex.
This complexity arises not only from the underlying algorithms but also from the challenges associated with the runtime integration.
In our approach, we distinguish between two phases of code generation: software code generation and integration code generation. This subsection focuses on the first phase.

We consider each subcomponent described in the event chain as a standalone piece of software that can be developed and tested independently before integration into the complete system. The event chain description of a subcomponent includes its behavior and implementation logic, both derived from the system requirements. This information is provided to the LLM through a structured prompt to guide the generation of code for the subcomponents. An example of such a prompt is shown in Appendix~\ref{lst:prompt_code}.

The prompt specifies that each subcomponent should be implemented as a standalone Python class. The resulting class is expected to process input signals directly from the middleware and produce outputs in a predefined structure, such as a Python dictionary. 
The primary functionality of the subcomponent should be encapsulated within an execute() method, which serves as the main entry point for its operation.
If a subcomponent is stateful, its internal state must be fully managed within the class. 
This design enables a static, model-based integration strategy to connect subcomponents within the overall vehicle software system (Section~\ref{sec:model-code-gen}).

\section{Model-based Formal Approach}
The model-based approaches introduced in this work aim to generate integration-related code, allowing different components generated by LLMs (Section~\ref{sec:LLM-code-gen}) to be integrated into the system and function together. At the same time, model-based method also enables formal system analysis, either through automated tools or by human engineers. 
\subsection{Meta model and instance model}\label{sec:model}
Models are fundamental elements in any model-based methodology.
According to the modeling standard, the Object Management Group's Meta-Object Facility (OMG MOF) \cite{OMG2016}, models are categorized to meta models and instance models based on their level of abstraction.
The meta model defines the abstract information of a group of systems. It serves as the modeling syntax or schema for constructing corresponding instance models.
In contrast, an instance model captures detailed information about a specific system and adheres to the structure defined by its corresponding meta model.
Widely adopted modeling languages and tools such as \ac{UML}~\cite{pilone2005uml}, \ac{SysML}~\cite{friedenthal2014practical} and \ac{EMF}~\cite{Steinberg2008} are compatible to this standard.

\begin{figure}[t]
	\centering
	\includegraphics[width=\linewidth]{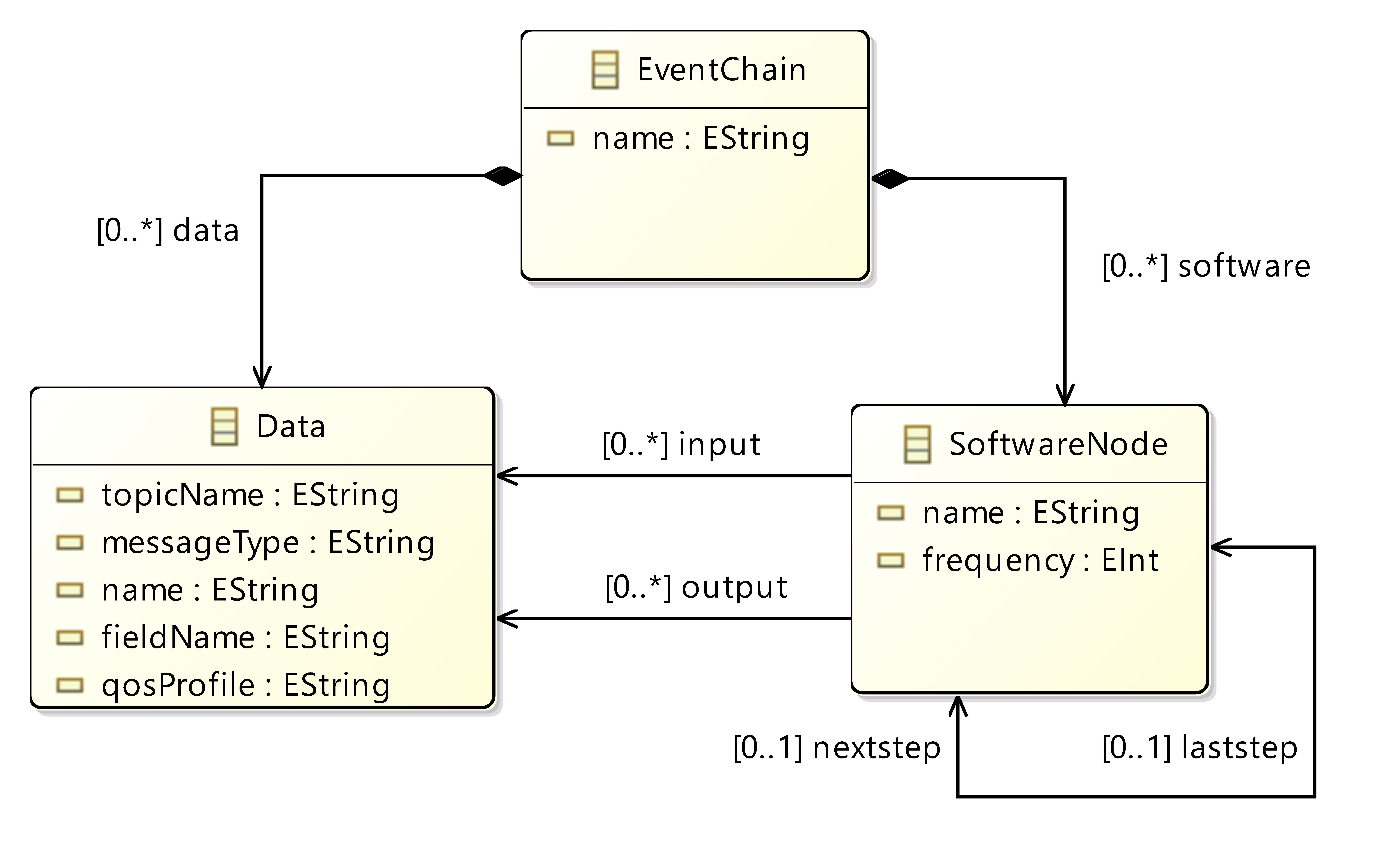}
	\caption{An example meta model for event chain}
	\label{fig:meta-model}
\end{figure}

To formally model event chains, we developed an example meta-model using EMF, as illustrated in Figure~\ref{fig:meta-model}. In our meta model, an event chain may consist of multiple software components and associated data elements. These data elements represent input and output signals of the software components. Attributes such as execution frequency, data topics, and message types are specified. These details are essential for enabling model-based code generation. Additional attributes can also be defined in the meta model for enabling further system analysis. However, comprehensive modeling strategies for enabling model-based system analysis will not be further discussed in this paper.

\begin{figure}[t]
	\centering
	\includegraphics[width=\linewidth]{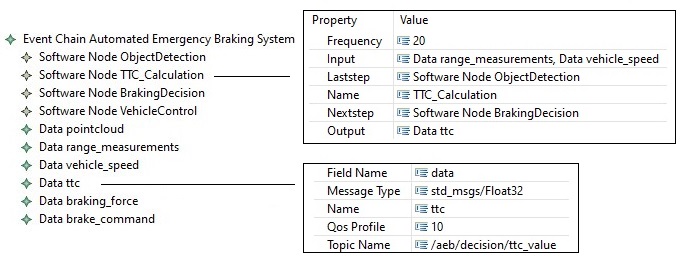}
	\caption{Generated instance model for the AEB case study}
	\label{fig:instance-model}
\end{figure}

In this work, we employ LLMs to automatically generate instance models based on our defined meta-model (Section~\ref{sec:model-gen}). An example of such a generated instance model, depicting a basic AEB system , is shown in Figure~\ref{fig:instance-model}. This AEB system serves as case study scenario in Section~\ref{sec:evaluation}.

\subsection{Constraints and Model Validation}\label{sec:OCL}
In the model-based domain, constraints are used to express rules that an instance model should obey but which cannot be defined in the meta model.
In this work, we utilize LLMs to generate OCL constraints for requirements, specifications, and design decisions (primarily non-functional related) that can be used for model validation (Section~\ref{sec:model-gen}).

For software described as an event chain, typical design constraints may include: 1. Each software node must have at least one input data and at least one output data. 2. The frequency of the software node in the next step must be higher than or equal to the frequency of the current software node.
Examples of generated OCL constraints are as follows:
\begin{lstlisting}[language=OCL]
context SoftwareNode
  inv HasInputAndOutputData:
    self.input->notEmpty() and self.output->notEmpty()
  inv NextstepFrequencyEqualOrHigher:
    self.nextstep->notEmpty() implies (self.nextstep.
    frequency >= self.frequency)
\end{lstlisting}

Using the Eclipse OCL Plugin or other solver-based formal methods mentioned in~\cite{pan2024sdv}, model validation can be performed and unsatisfied constraints with respect to a specific instance model can be detected.

\subsection{Model-based Code Generation}\label{sec:model-code-gen}
We employ model-to-text generation techniques to produce middleware-specific code that integrates individual subcomponents into the vehicle runtime.
For models adhering to the OMG MOF standard, the OMG MOF Model to Text Transformation Language~(MOFM2T) specification defines a standard method for generating textual artifacts from MOF-based models. We use Acceleo~\cite{acceleo}, a model-to-text generation tool, to generate code from the information contained within instance models.

The code generation process takes three inputs: the meta-model, the instance model, and a code generation template. In Acceleo, templates are defined using the Acceleo Query Language (AQL).
As discussed previously in Section~\ref{sec:LLM-code-gen}, the core algorithms of subcomponents are generated by LLMs and are middleware-independent. Here, model-based generation serves the purpose of bridging these LLM-generated modules with the system runtime.
To integrate subcomponents into different systems with varying middleware architectures, separate generation templates must be defined for each middleware.

An example code template for generating ROS-specific code to integrate LLM-generated software components in our demonstration scenario is presented in Appendix~\ref{lst:acceleo}. The red text indicates AQL operations, including the entry point of generation. 
The green text represents comments, which also provide clarification and refer to the main Acceleo class implementation.
The blue text denotes AQL queries. These queries are written using meta-information (e.g., class and property definitions) from the meta model and are used to retrieve specific objects defined in the instance model.
The rest of texts are predefined static codes constructing the class of a ROS node.
The initial lines of the template specify generation configuration, including the entry point, which in our case is the \textit{EventChain} class. We then iterate over each \textit{SoftwareNode} within the event chain and retrieve required attributes through AQL queries, generating a corresponding Python script for each as an individual ROS node that can be directly deployed into ROS runtime. The node structure is predefined within the template.
Signal publishing, subscription mechanisms, and submodule interactions are generated based on the data preserved in the instance model. In general, each generated ROS node receives signals, passes them to the corresponding LLM-generated core module, obtains the result, and then publishes it for subsequent processing.

%% file: sections/04_evaluation.tex
In this section, we demonstrate the feasibility of the proposed automotive software development approach using a basic AEB scenario.

In the implementation, we utilized Eclipse IDE 2022-06 to create the event chain meta-model (Fig.~\ref{fig:meta-model}) and to present the generated instance model (Fig.~\ref{fig:instance-model}). The OCL All-In-One SDK (version 6.17.1.v20220309-0840) was used for model validation.
The Acceleo tool (version 3.7.11.202102190929) served as the engine for model-to-text code generation.

As LLMs, we used and compared GPT-4o, Gemini 2.5 Pro and Llama 3.3 70B for their performance in generating models and codes. The commercial LLMs, GPT-4o and Gemini 2.5 Pro, were accessed via their official APIs. The open source Llama 3.3 70B was executed through the Groq platform. The experiment of each LLM was conducted 5 times. Each LLM was evaluated across five independent runs. The comparison results are presented in Section~\ref{sec:comparison}.

We further set up a CARLA simulation to verify the generated code. We utilize CARLA version 0.9.15, and choose Town 01 as the map. A straight lane before the traffic light replicates a typical scenario that could trigger the brake system. A stopped vehicle is spawned before the red light, and our ego vehicle should stop before crashing. A Carla-ros-bridge has been built to exchange sensor and actuator information between ROS and the CARLA environment. Specifically, the ROS2 Foxy version of the bridge is used for improved real-time performance. The simulation runs on a PC equipped with an Intel Core i7-6850K CPU, 32 GB RAM, and two Nvidia GTX 1080 GPUs.

\subsection{AEB Case Study}\label{sec:AEB}
This subsection presents one successful end-to-end execution using GPT-4o.
The following requirements were used for generating the AEB software:
\begin{itemize}
    \item The system shall receive distance and relative speed data from simulated or physical lidar sensors.
    \item The system shall calculate Time-To-Collision (TTC) using object distance and relative speed.
    \item The system shall signal an emergency braking condition when TTC falls below 1.0 seconds.
    \item The system shall determine brake force based on TTC thresholds: Full brake if TTC $<$ 1.0s, Partial brake if 1.0s $\leq$ TTC $<$ 2.0s, No brake if TTC $\geq$ 2.0s
    \item The system shall output a normalized brake force (0.0 to 1.0).
    \item The system shall command braking force to the actuator based on the Braking Force Command output.
\end{itemize}

As a preliminary step, we constructed two ROS nodes to serve as existing software components.
The \textit{ObjectDetection} node integrates the open-source \textit{pointcloud\_to\_laserscan} package~\cite{pointcloudtolaser2025}, which converts LiDAR point cloud data from CARLA into \textit{LaserScan} messages containing the shortest distances to nearby obstacles.
The \textit{VehicleControl} node receives control commands and translates them into vehicle control messages for braking and other maneuvers.
Additionally, two dummy software components were included for simulation purposes. 
Input information about the existing ROS topics and their message definitions was also collected.

\begin{figure}[t]
	\centering
	\includegraphics[width=\linewidth]{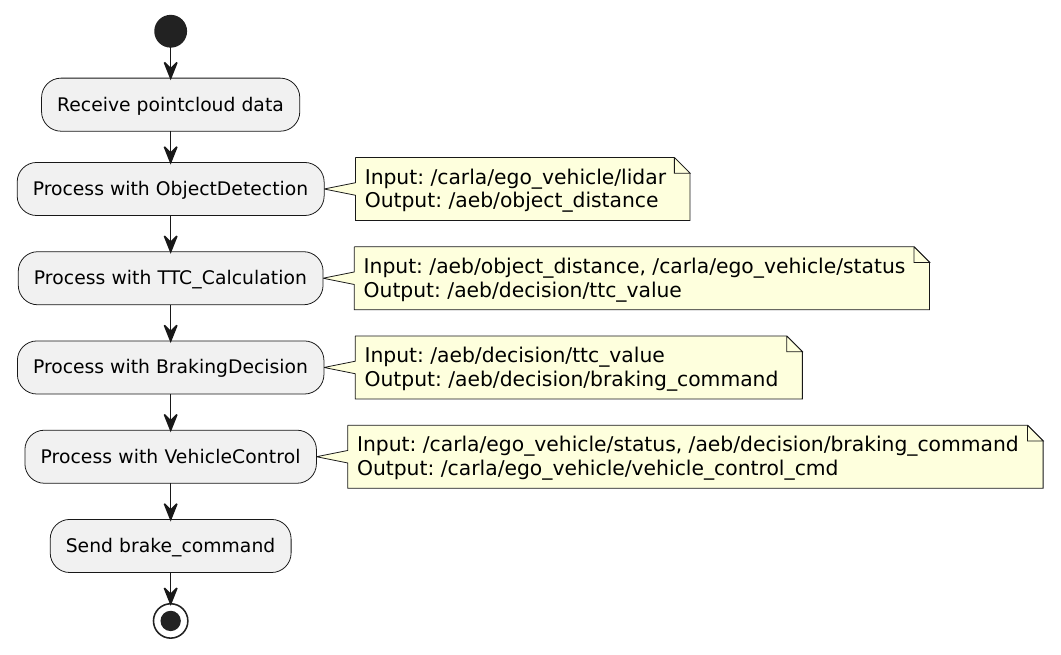}
	\caption{Illustration of generated AEB event chain}
	\label{fig:event_chain}
\end{figure}

The event chain generation follows the process described in Section~\ref{sec:eventgen}. The resulting AEB event chain includes four software components. To illustrate the outcome, we provide a manually created activity diagram using PlantUML (Fig.~\ref{fig:event_chain}). The event chain begins with the \textit{ObjectDetection} node, which receives LiDAR point cloud data and outputs a \textit{LaserScan} message with object distances. This message, along with current vehicle status, is passed to the generated \textit{TTC\_Calculation} node to compute the shortest Time-To-Collision. The \textit{Braking\_Decision} node then determines the required braking force, and the \textit{Carla\_Vehicle\_Control} node sends the corresponding control commands to the vehicle.

By comparing the event chain description with the list of existing components, we identified that \textit{TTC\_Calculation} and \textit{BrakingDecision} were not yet implemented and needed to be generated.
We passed their descriptions to the prompt in Appendix~\ref{lst:prompt_code} and generated the function code as individual Python classes for each component.

\begin{figure}[t]
	\centering
	\includegraphics[width=0.45\textwidth]{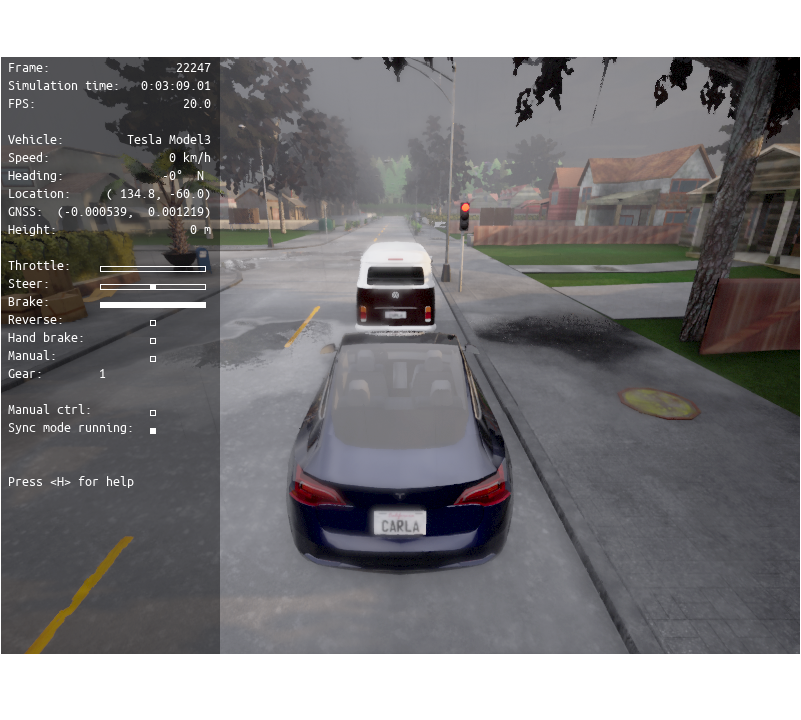}
	\caption{Functioning in CARLA simulator}\label{fig:carla}
\end{figure}

Simultaneously, the entire event chain was modeled as a formal EMF instance (Fig.~\ref{fig:instance-model}) based on the event chain meta-model (Fig.~\ref{fig:meta-model}) for further inspection and validation.
The instance model successfully passed the early-phase validation of the OCL constraints presented in Section~\ref{sec:OCL} using Eclipse OCL.
Afterwards, the instance model was  utilized via model-based code generation (Section~\ref{sec:model-code-gen}) to generate integration-related code that integrates the function code of each component as ROS nodes.

The AEB software was executed in the predefined simulation scenario.
The generated AEB module was successfully triggered, and the ego vehicle responded by braking in time before reaching the stationary vehicle (Fig.~\ref{fig:carla}).
This basic AEB system demonstrates the feasibility of our development approach.

\subsection{Comparison of LLMs} \label{sec:comparison}
To evaluate the performance of different LLMs in our agentic workflow, we measured their ability to generate both event chain models and functional code components. Each LLM was tested in five complete development runs.

We expect the generated artifacts to be accurate both semantically (with respect to the target modeling and programming languages) and syntactically (by conforming to the given system requirements). The evaluation results are summarized in Table~\ref{table: comparison}. To be considered successful, a single execution must result in both a valid event chain model and functional code, and the resulting AEB system must perform correctly within the simulation environment.

Among the evaluated models, GPT-4o achieved the highest overall success rate (60\%), showing strongest performance in event chain modeling (80\%). Gemini 2.5 Pro produced the most reliable function code (100\%) but exhibited lower overall performance due to difficulties in generating valid event chain models. In contrast, Llama 3.3 70B showed limited effectiveness in both tasks, with no successful end-to-end executions.

These results highlight the complementary strengths of different LLMs, suggesting that a multi-agent workflow, where distinct LLMs are assigned to specific sub-tasks, could enhance the robustness and reliability of the overall development pipeline.

\begin{table}[h]
	\caption{Accuracy of generated artifacts}
	\centering
	\label{table: comparison}
	\sisetup{group-digits=false,table-number-alignment=right,table-parse-only=true}%
	\begin{tabular}{lccc}
		\toprule
		LLM & eventchain model & function code & overall success\\
		\midrule
		GPT-4o & \textbf{80\%} & 60\% & \textbf{60\%} \\
		Gemini 2.5 Pro & 40\% & \textbf{100\%} & 40\%\\
		Llama 3.3 70B & 20\% & 20\% & 0\%\\
		\bottomrule
	\end{tabular}
\end{table}

\subsection{Further discussion}
The proposed approach combines model-based engineering with generative AI to support the development of automotive software systems. The model-based component ensures reliable system integration by addressing non-functional aspects such as safety requirements and design constraints. Meanwhile, LLMs are employed to enhance automation in development activities, including functional requirement analysis, code generation, etc.
Due to the statistical nature of LLMs, the quality correctness of generated functional behavior cannot be guaranteed. Rigorous testing remains essential to validate the generated artifacts.

In conventional software development, testing, including unit testing, integration testing, and system testing, is critical for ensuring software quality. Within our workflow, integration code is generated using a correctness-by-construction approach, leveraging formal model definitions and predefined code templates. This eliminates the need for manual integration testing, as structural correctness is inherently assured by the model-based code generation process.
However, because the functional code is generated by LLMs, its behavior must still be validated through dedicated testing procedures. Consequently, in the proposed workflow, testing efforts can be concentrated on unit testing of LLM-generated components and system-level testing of the overall vehicle behavior.

%% file: sections/05_conclusion.tex
In this work, we introduced an agentic approach that combines the emerging capabilities of generative AI with model-driven formal methods to automate the development of automotive software.
The approach leverages \acp{LLM} to analyze requirements, construct the overall software design as an event chain model, and generate platform-independent software function code as standalone classes.
The formal event chain model serves as a basis for further (automated or human-centered) model-based system analysis and is also used to generate integration code, enabling the seamless connection of independent software components to the broader vehicle system via middleware.
The resulting system can be deployed and evaluated in a simulation environment, such as CARLA.
Simulation data can then be fed back into the event chain model to support further analysis, particularly in relation to non-functional requirements.
We demonstrated the feasibility of this approach through a basic AEB scenario.

Our future work will focus on extending the evaluation to larger event chains involving a greater number of software components. 
This will allow us to assess the scalability and robustness of the proposed approach in more complex automotive scenarios.
In parallel, we plan to investigate automated test generation techniques for validating LLM generated software.

%% file: ref.bib
@manual{OMG2016,
	title  = {Meta Object Facility},
	author = {OMG},
	version  = {2.5.1},
	year   = {2016}
}

@misc{muenzenberger2021event,
  title        = {Event-Chain-Centric Architecture Design of Driver Assistance Systems},
  author       = {Heckmann, Frieder and M\"unzenberger, Ralf},
  year         = {2021},
  howpublished = {\url{http://inchron.com/wp-content/uploads/2021/12/ESE21\_Muenzenberger\_Heckmann.pdf}},
  note         = {Accessed: 2025-04-07}
}

@article{phatale2024,
    author = {Phatale, Amey and Kaushik, Arpita},
    year = {2024},
    month = {11},
    pages = {1-5},
    title = {Generative AI Adoption in Automotive Vehicle Technology: Case Study of Custom GPT},
    volume = {3},
    journal = {Journal of Artificial Intelligence \& Cloud Computing},
    doi = {10.47363/JAICC/2024(3)400}
}

@inproceedings{padma2020,
author = {Iyenghar, Padma and Huning, Lars and Pulvermueller, Elke},
year = {2020},
pages = {477-489},
booktitle = {15th International Conference on Evaluation of Novel Approaches to Software Engineering},
title = {Automated End-to-End Timing Analysis of AUTOSAR-based Causal Event Chains},
doi = {10.5220/0009512904770489}
}

@InProceedings{patil2024,
author="Patil, Minal Suresh
and Ung, Gustav
and Nyberg, Mattias",
editor="Steffen, Bernhard",
title="Towards Specification-Driven LLM-Based Generation of Embedded Automotive Software",
booktitle="Bridging the Gap Between AI and Reality",
year="2025",
publisher="Springer Nature Switzerland",
address="Cham",
pages="125--144",
}

@misc{nouri2025,
author = {Nouri, Ali and Andersson, Johan and Hornig, Kailash and Fei, Zhennan and Knabe, Emil and Sivencrona, Hakan and Cabrero-Daniel, Beatriz and Berger, Christian},
year = {2025},
pages = {1-11},
title = {On Simulation-Guided LLM-based Code Generation for Safe Autonomous Driving Software},
doi = {10.48550/arXiv.2504.02141}
}

@misc{abdalla2024,
  author       = {Abdelrahman Abdalla and Harsh Pandey and Behzad Shomali and Joschka Schaub and Arne Müller and Markus Eisenbarth and Jakob Andert},
  title        = {Generative Artificial Intelligence for Model-Based Graphical Programming in Automotive Function Development},
  year         = {2024},
  month        = {November 10},
  note         = {Available at SSRN: \url{https://ssrn.com/abstract=5153452} or \url{http://dx.doi.org/10.2139/ssrn.5153452}},
}

@misc{pan2025model,
  title={LLM-enabled Instance Model Generation},
  author={Fengjunjie Pan and Nenad Petrovic and Vahid Zolfaghari and Long Wen and Alois Knoll},
  year={2025},
  eprint={2503.22587},
  archivePrefix={arXiv},
  primaryClass={cs.SE},
  doi={10.48550/arXiv.2503.22587}
}

@inproceedings{holtmann2021,
author = {Holtmann, Jörg and Meyer, Jan and Meyer, Matthias},
year = {2011},
month = {02},
pages = {79-88},
title = {A Seamless Model-Based Development Process for Automotive Systems},
}

@Inbook{VinothKannan2021,
author="Vinoth Kannan, K.",
editor="Kathiresh, M.
and Neelaveni, R.",
title="Model-Based Automotive Software Development",
bookTitle="Automotive Embedded Systems: Key Technologies, Innovations, and Applications",
year="2021",
publisher="Springer International Publishing",
address="Cham",
pages="71-87",
isbn="978-3-030-59897-6",
doi="10.1007/978-3-030-59897-6_4",
}

@Manual{OMG2014,
	title  = {Object Constraint Language Version 2.4},
	author = {OMG},
	month  = feb,
	year   = {2014},
}

@ARTICLE{pan2024sdv,
  author={Pan, Fengjunjie and Rickert, Markus and Betz, Tobias and Wen, Long and Lin, Jianjie and Petrovic, Nenad and Lienkamp, Markus and Knoll, Alois},
  journal={IEEE Access}, 
  title={Toward Software-Defined Vehicles: From Model-Based Engineering to Virtualization-Based Deployment}, 
  year={2024},
  volume={12},
  number={},
  pages={192127-192145},
}

@misc{petrovic2025meta,
  title={LLM-based Iterative Approach to Metamodeling in Automotive},
  author={Nenad Petrovic and Fengjunjie Pan and Vahid Zolfaghari and Alois Knoll},
  year={2025},
  eprint={2503.05449},
  archivePrefix={arXiv},
  primaryClass={cs.SE},
  doi={10.48550/arXiv.2503.05449}
}

@INPROCEEDINGS{pan2024ocl,
  author={Pan, Fengjunjie and Zolfaghari, Vahid and Wen, Long and Petrovic, Nenad and Lin, Jianjie and Knoll, Alois},
  booktitle={2024 IEEE International Symposium on Systems Engineering (ISSE)}, 
  title={Generative AI for OCL Constraint Generation: Dataset Collection and LLM Fine-tuning}, 
  year={2024},
  volume={},
  number={},
  pages={1-8},
  doi={10.1109/ISSE63315.2024.10741141}}

@INPROCEEDINGS{petrovic2024-3,
  author={Petrovic, Nenad and Lebioda, Krzysztof and Zolfaghari, Vahid and Schamschurko, André and Kirchner, Sven and Purschke, Nils and Pan, Fengjunjie and Knoll, Alois},
  booktitle={2024 2nd International Conference on Foundation and Large Language Models (FLLM)}, 
  title={LLM-Driven Testing for Autonomous Driving Scenarios}, 
  year={2024},
  volume={},
  number={},
  pages={173-178},
  doi={10.1109/FLLM63129.2024.10852505}}

@misc{matarazzo2025survey,
  author       = {Andrea Matarazzo and Riccardo Torlone},
  title        = {A Survey on Large Language Models with some Insights on their Capabilities and Limitations},
  year         = {2025},
  eprint       = {2501.04040v1},
  archivePrefix= {arXiv},
  primaryClass = {cs.CL},
  url          = {https://arxiv.org/abs/2501.04040v1}
}

@inproceedings{dosovitskiy2017carla,
  title={CARLA: An open urban driving simulator},
  author={Dosovitskiy, Alexey and Ros, German and Codevilla, Felipe and Lopez, Antonio and Koltun, Vladlen},
  booktitle={Conference on robot learning},
  pages={1--16},
  year={2017},
  organization={PMLR}
}

@Misc{pointcloudtolaser2025,
  title     = {ROS 2 pointcloud $<$-$>$ laserscan converters},
 howpublished = {\url{https://github.com/ros-perception/pointcloud\_to\_laserscan}},
 note         = {Accessed: 2025-04-18},
}

@Misc{mckinsey,
  title     = {The case for an end-to-end automotive-software platform},
  howpublished = {\url{https://www.mckinsey.com/industries/automotive-and-assembly/our-insights/the-case-for-an-end-to-end-automotive-software-platform}},
  note         = {Accessed: 2025-03-28},
}

@book{pilone2005uml,
	title={UML 2.0 in a Nutshell},
	author={Pilone, Dan and Pitman, Neil},
	year={2005},
	publisher={" O'Reilly Media, Inc."}
}

@book{friedenthal2014practical,
	title={A practical guide to SysML: the systems modeling language},
	author={Friedenthal, Sanford and Moore, Alan and Steiner, Rick},
	year={2014},
	publisher={Morgan Kaufmann}
}

@book{Steinberg2008,
	title={EMF: eclipse modeling framework},
	author={Steinberg, Dave and Budinsky, Frank and Merks, Ed and Paternostro, Marcelo},
	year={2008},
	publisher={Pearson Education}
}

@misc{acceleo,
  author       = {OBEO},
  title        = {GENERATE ANYTHING FROM ANY EMF MODEL},
  howpublished = {\url{https://eclipse.dev/acceleo/}},
  note         = {Accessed: 2024-03-25}
}
